\documentclass[aps,prd,twocolumn,numerical,floatfix,nofootinbib]{revtex4-2}
\usepackage{physics}
\usepackage{lipsum}
\makeatletter
\def\ps@pprintTitle{%
 \let\@oddhead\@empty
 \let\@evenhead\@empty
 \def\@oddfoot{}%
 \let\@evenfoot\@oddfoot}
\makeatother

\usepackage{float,ulem}
\usepackage{graphicx,epsfig,epstopdf,amsmath,amssymb}
\usepackage{xcolor}
\usepackage[colorlinks=true,linkcolor=blue,citecolor=blue,urlcolor=
blue]{hyperref}

\newcommand{\be}{\begin{equation}}
\newcommand{\ee}{\end{equation}}
\newcommand{\ba}{\begin{eqnarray}}
\newcommand{\ea}{\end{eqnarray}}
\newcommand{\nn}{\nonumber\\}
\def\k{\boldsymbol k}
\def\v{\boldsymbol {\bf{v}}}
\def\om{\omega}
\def\gm{\gamma}
\usepackage{caption}
\usepackage{subcaption}

\begin{document}
\title{Causality Criteria from Stability Analysis at Ultra-High Boost}

\author{Shuvayu Roy}
\email{shuvayu.roy@niser.ac.in}

\author{Sukanya Mitra}
\email{sukanya.mitra@niser.ac.in}
%\affiliation{School of Physical Sciences, National Institute of Science Education and Research, An OCC of Homi Bhabha Nuclear Institute, Jatni-752050, India}

\address{School of Physical Sciences, National Institute of Science Education and Research, An OCC of Homi Bhabha National Institute, Jatni-752050, India.}

\begin{abstract}
In this work, we have exclusively employed the linear stability analysis at ultra-high boost on two well-known stable-causal theories - second-order MIS and first-order BDNK, to identify the region of parameter space over which they are frame-invariantly stable and obey causal signal propagation. It has been shown for these two theories that at near-luminal boost, stability criteria alone can provide the causality constraints on transport coefficients, which are identical to the asymptotic causality conditions, without actually going to the asymptotic limit of the theories. The analysis indicates an alternative approach to derive the causality constraints, which is more appropriate for low-energy effective theories like relativistic hydrodynamics.
\end{abstract}

\maketitle

\section{Introduction}
Hydrodynamics is an effective theory that describes the dynamical evolution of the conserved quantities (the state variables of the system) at low energy, long wavelength limit \cite{Landau} and has long served as a powerful tool to study the collective behaviour of a system \cite{Baier:2007ix,Huovinen:2006jp}. Its chronological development began with ideal hydrodynamics where the fluid is in its equilibrium state, then subsequently went on to include dissipative corrections for out of equilibrium scenario. These corrections are formulated by a systematic build up of gradients on the fundamental hydrodynamic fields \cite{Romatschke:2009im}. For each order of hydrodynamic gradient expansion, the transport coefficients from the underlying microscopic theory enter the hydrodynamic evolution equations as a dynamical
input of the system interaction.
 
 Based on these foundations, it is possible to derive a number of alternative hydrodynamic theories following different approaches. However, a reliable, pathology free theory needs to guarantee two major
 benchmark criteria. First, the signal propagation predicted by the equations of motion of the theory must be subluminal and second, its equilibrium state should be stable against fluctuations i.e., the fluctuations must not grow indefinitely with time. Now, in a number of works it has been established that the group velocity of the propagating mode exceeding the speed of light for some frequency range does not violate causality, as long as it is subluminal at the infinite frequency (wavenumber) limit \cite{Fox:1970cu,Pu:2009fj}. This necessary condition for causality is called the asymptotic causality condition which has been widely used to check the causal validity of a hydrodynamic theory \cite{Denicol:2008ha,Brito:2020nou,Mitra:2021ubx}. But the conceptual anomaly with this approach is that the hydrodynamic gradient expansion has been tested to be a divergent 
 series with factorial growth of large order corrections indicating a zero radius of convergence \cite{Heller:2013fn,Heller:2015dha}. Given the situation, an alternate definition of causality is imperative.   
 On the other hand, the stability of a relativistic system has been known to behave distinctly depending upon the observer's frame of reference \cite{Hiscock:1985zz}. This issue has been recently addressed in \cite{Gavassino:2021owo,Gavassino:2023myj}, where it has been argued that frame-invariant stability is possible only if the theory respects causality. The objective of the current work is to employ the frame-invariance of the stability property of a theory to establish its causality constraints. The non-triviality again comes from the fact that checking linear stability at arbitrary reference frames to identify the invariantly stable parameter space can be a cumbersome job. In this work for two well-known stable-causal theories, we have demonstrated that the linear stability analysis in a reference frame boosted to a near luminal speed can alone provide the stability invariant parameter space at the spatially homogeneous limit of the theory and hence can be used to determine the causal domain of the theory as well. In \cite{Biswas:2022hiv}, this identification has been observed from a kinetic theory derivation of a stable-causal first-order theory. Here, we show that one can solely use the low-wavenumber stability analysis to produce the exact results of asymptotic causality in the MIS and BDNK theories. The analysis presented here serves as a case study of two most well-known stable-causal theories to show that the causality of a theory can be probed without departing from the small-$k$ domain. Since relativistic hydrodynamics is a low-energy effective theory, hence we believe this approach provides us with a more appropriate definition of causality.

\section{Basic setup}

 In this work, hydrodynamic stability has been analyzed in a generalized Lorentz frame with an arbitrary boost velocity for both second-order M\"{u}ller-Israel-Stewart (MIS) theory \cite{Israel:1976tn,Israel:1979wp,Muller:1967zza}, and the recently proposed first-order stable-causal (BDNK) theory \cite{Bemfica:2017wps,Bemfica:2019knx,Kovtun:2019hdm,Hoult:2020eho,Rocha:2022ind}. We linearize the conservation equations for small perturbations of fluid variables around their hydrostatic equilibrium,
 $\psi(t,x)=\psi_0+\delta\psi(t,x)$, with the fluctuations expressed in the plane wave solutions via a Fourier transformation $\delta\psi(t,x)\rightarrow e^{i(kx-\omega t)} \delta\psi(\omega,k)$, 
 (subscript $0$ indicates global equilibrium). The background fluid is considered to be boosted along the x-axis with a constant velocity $\textbf{v}$, $u^{\mu}_0=\gamma(1,\textbf{v},0,0)$ with 
 $\gamma=1/\sqrt{1-\textbf{v}^2}$. The corresponding velocity fluctuation is $\delta u^{\mu}=(\gamma \textbf{v} \delta u^x,\gamma\delta u^{x},\delta u^y,\delta u^z)$ which gives $u^{\mu}_0\delta u_{\mu}=0$ 
 to maintain the velocity normalization. In the following analysis, we present the leading order stability analysis (at $k\rightarrow 0$ limit) for both the theories at conformal, charge less limit. 

\section{Conventions and notations}
Throughout the manuscript, we have used natural unit~($\hbar = c = k_{B} = 1 $) and flat space-time with mostly positive metric signature $\eta^{\mu\nu} = \text{diag}\left(-1,1,1,1\right)$. 
The used notations read, $D\equiv u^{\mu}\partial_{\mu}$, $\nabla^{\mu}=\Delta^{\mu\nu}\partial_{\nu}$, 
$\sigma^{\mu\nu}=\Delta^{\mu\nu}_{\alpha\beta} \partial^{\alpha}u^{\beta}$ with $\Delta^{\mu\nu\alpha\beta}=\frac{1}{2}\Delta^{\mu\alpha}\Delta^{\nu\beta}+\frac{1}{2}\Delta^{\mu\beta}\Delta^{\nu\alpha}
-\frac{1}{3}\Delta^{\mu\nu}\Delta^{\alpha\beta}$ and
$\Delta^{\mu\nu}=\eta^{\mu\nu}+u^{\mu}u^{\nu}$, $\epsilon\equiv$ energy density, $P\equiv$ pressure, $u^{\mu}\equiv$ hydrodynamic four-velocity, 
$\tau_{\pi}\equiv$ relaxation time of shear-viscous flow, $\eta \equiv$ shear viscous coefficient,
${\mathcal{E}},\theta$ are first order field correction coefficients of BDNK theory. From the constraints of the second law of thermodynamics, $\eta$ should always be a positive number
\cite{Weinberg:1971mx}. The scaling notation $\tilde{x}$ denotes $x/(\epsilon_0+P_0)$. 

\section{Identifying stability invariant parameter space from ultra-high boost}

 First, we discuss the case of MIS theory where the energy-momentum tensor takes the form, $T^{\mu\nu}=\epsilon u^{\mu}u^{\nu}+P\Delta^{\mu\nu}+\pi^{\mu\nu}$. The conservation of energy-momentum tensor $\partial_{\mu}T^{\mu\nu}=0$ and the relaxation equation of shear viscous flow $\pi^{\mu\nu}=-\tau_{\pi}\Delta^{\mu\nu}_{\alpha\beta}D\pi^{\alpha\beta}-2\eta \sigma^{\mu\nu}$ together give us the equations of motion to be linearized. In the transverse or shear channel, the leading term of the frequency ($\omega$) solution in wavenumber $k$-expansion is a single non-hydro non-propagating mode, $\omega^{\perp}_{\textrm{MIS}}=-i/\gamma(\tau_{\pi}-\tilde{\eta}\textbf{v}^2)$. 
 Now the demand that stability requires the imaginary part of the frequency to be negative renders the stability criteria $\tau_{\pi}/\tilde{\eta}>\textbf{v}^2$ \cite{Pu:2009fj}. For sound channel, the leading order single non-propagating mode turns out to be, $\omega^{\parallel}_{\textrm{MIS}}=-i(1-\frac{\textbf{v}^2}{3})/\gamma[\tau_{\pi}(1-\frac{\textbf{v}^2}{3})-\frac{4\tilde{\eta}}{3}\textbf{v}^2]$. For the range of boost velocity $0\leq \textbf{v} < 1$, the stability condition becomes, $\tau_{\pi}/\tilde{\eta}>\frac{4}{3}\textbf{v}^2/(1-\frac{\textbf{v}^2}{3})$. In both the channels, the right-hand sides of the inequalities for $\tau_{\pi}/\tilde{\eta}$ are monotonically increasing functions of $\textbf{v}$ within the mentioned range that allow only positive values of $\tau_{\pi}$ and give the strictest bound for $\textbf{v}\rightarrow1$. So we infer that the allowed parameter space over the transport coefficients 
 $\eta$ and $\tau_{\pi}$ set by stability criteria at the spatially homogeneous limit ($k\rightarrow0$) for any boost velocity $\textbf{v}$, is always a subset of the same for any lower value of $\textbf{v}$. Hence, we conclude here that the $\textbf{v}\rightarrow1$ bound ($\tau_{\pi}>\tilde{\eta}$ for shear channel and $\tau_{\pi}>2\tilde{\eta}$ for sound channel) provides the necessary and sufficient region in the parameter space where the system is stable at the spatially homogeneous limit for all reference
 frames ($0\leq \textbf{v}< 1$). So here we see that for the MIS theory, checking stability alone in a reference frame with ultra-high boost $(\textbf{v}\rightarrow 1)$ is sufficient to identify the frame-invariantly stable parameter space at $k\rightarrow0$ limit.
 
\begin{figure}[h!]
    \centering
    \includegraphics[width=0.42\textwidth]{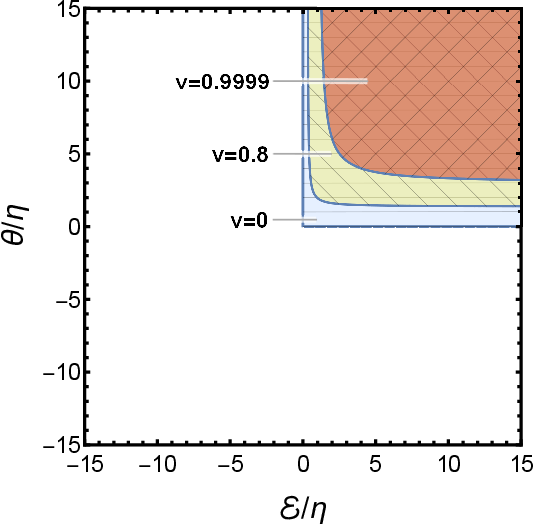}
    \caption{Linearly stable parameter space for BDNK sound channel for different $\textbf{v}$ values.}
    \label{fig:my_label}
\end{figure}

 Next, we discuss the case of BDNK theory for which the energy-momentum tensor takes the form, 
 $T^{\mu\nu}=(\epsilon +\epsilon_1)u^{\mu}u^{\nu}+(P+ P_1)\Delta^{\mu\nu}+(u^{\mu}W^{\nu}+u^{\nu}W^{\mu})+\pi^{\mu\nu}$, with the first order dissipative field corrections, 
 $\epsilon_1 ={\mathcal{E}}\frac{D\epsilon}{\epsilon_0+P_0}+{\mathcal{E}}(\partial\cdot u),  P_1 =\frac{{\mathcal{E}}}{3}\frac{D\epsilon}{\epsilon_0+P_0}+\frac{{\mathcal{E}}}{3}(\partial\cdot u),
 W^{\mu}=\theta[\frac{\nabla^{\mu}T}{T}+Du^{\mu}]$ and $\pi^{\mu\nu}=-2\eta\sigma^{\mu\nu}$. The shear channel analysis is identical to that of MIS theory with the replacement $\tau_{\pi}=\theta/(\epsilon_0+P_0)$
 \cite{Kovtun:2019hdm}. However, the situation becomes significantly more mathematically involved in the sound channel. The leading order $\omega$ solution in $k$-expansion gives rise to the quadratic dispersion
 relation $a\omega^2+b\omega+c=0$, with $a=\gamma^2[\tilde{\cal{E}}\tilde{\theta}-\frac{2}{3}\tilde{\cal{E}}(2\tilde{\eta}+\tilde{\theta})\textbf{v}^2+\frac{1}{9}\tilde{\theta}(\tilde{\cal{E}}-4\tilde{\eta})\textbf{v}^4]$, $b=i\gamma[(\tilde{\cal{E}}+\tilde{\theta})-\frac{1}{3}(\tilde{\theta}+\tilde{\cal{E}}+4\tilde{\eta})\textbf{v}^2]$ and $c=(\textbf{v}^2/3-1)$. This dispersion polynomial gives rise to two non-propagating, non-hydro modes whose stability has been analyzed using the Routh-Hurtwitz (R-H) stability test \cite{Korn}. The stability criteria constrain the parameter space for BDNK sound channel through the two following inequalities, 
 \begin{align}
 &{\cal{E}}\theta\left(1-\frac{\textbf{v}^2}{3}\right)^2-\frac{4}{3}\eta\textbf{v}^2\left({\cal{E}}+\frac{\textbf{v}^2}{3}\theta\right)>0~,
 \label{BDNK-snd-stab1}\\
 &\left({\cal{E}+\theta}\right)\left(1-\frac{\textbf{v}^2}{3}\right)-\frac{4}{3} \eta \textbf{v}^2 >0~.
 \label{BDNK-snd-stab2}
 \end{align}
 Eq.\eqref{BDNK-snd-stab1} and \eqref{BDNK-snd-stab2} together necessarily confine the parameter space within the region,
 \begin{align}
 \frac{\theta}{\eta}>\frac{4}{3}\frac{\textbf{v}^2}{\left(1-\textbf{v}^2/3\right)^2}~~,~~~~~\frac{\cal{E}}{\eta}>\frac{4}{9}\frac{\textbf{v}^4}{\left(1-\textbf{v}^2/3\right)^2}~.
 \label{BDNK-snd-stab3}
 \end{align}
 The right-hand sides of both the inequalities are monotonically increasing functions of $\textbf{v}$ which allow only positive values of $\cal{E}$ and $\theta$ with lower bounds ranging from $0$ to $\eta$ and $0$ 
 to $3\eta$ respectively as $\textbf{v}$ ranges from $0$ to $1$. Following these conditions, Fig.\ref{fig:my_label} shows that the parameter space where the theory is stable at $\textbf{v}\rightarrow 1$ is enclosed within the same for any lower value of $\textbf{v}$. So, identical to the situation of MIS theory, for BDNK theory as well, the stability condition at $\textbf{v}\rightarrow 1$, is a necessary and sufficient condition for stability to hold at the spatially homogeneous limit for all possible boost velocities $0\leq\textbf{v}<1$. 
 %The detailed dispersion polynomials for both theories are given in the Appendix.

 Given the above analysis for MIS and BDNK theories, we establish our first key finding here. For relativistic dissipative hydrodynamic theories like BDNK and MIS, performing stability analysis at ultra-high
 boost velocity ($\textbf{v}\rightarrow 1$) alone suffices to conclude the stability invariance of the theory. Stability analysis at any other boost velocity lacks this confirmation. The stable
 parameter space at $\textbf{v}\rightarrow 1$ is a necessary and sufficient region of the theory for stability invariance to hold at the spatially homogeneous limit. 

\section{Causality from stability analysis}

In this section, we will prove that only the stability criteria at $\textbf{v} \rightarrow 1$ limit is enough to provide the region of parameter space over which each of these two theories is causal. The idea is that, since it has been proven for theories like MIS and BDNK that the stability conditions at $\textbf{v} \rightarrow 1$ identify the region of parameter space where the system is frame invariantly stable, and since stability invariance requires the causality properties of the theory to be respected according to the arguments put forward in \cite{Gavassino:2021owo,Gavassino:2023myj}, hence the stability constraints at ultra-high boost automatically lead us to the causal region of the parameter space.
For MIS theory, the stability conditions at  $\textbf{v} \rightarrow 1$ limit for the shear and sound channels give us
$\frac{\tau _{\pi }}{\tilde{\eta }}>1$ and $\frac{\tau _{\pi }}{2 \tilde{\eta }}>1$ respectively. It can be shown that the expressions on the left-hand sides of the inequalities for both channels are functions of the square of respective asymptotic group velocities $v_g=\lim _{k\to \infty }\left|\frac{\partial  \text{Re} (\omega) }{\partial  k}\right|$, $(v_g^{2})^{\perp}=\tilde{\eta}/\tau_{\pi}$ and
$(v_g^{2})^{\parallel}=\frac{4\tilde{\eta}}{3\tau_{\pi}}+\frac{1}{3}$. These expressions for both the channels finally reduce to $0<v_g^2<1$, and therefore, the stability criteria at  $\textbf{v} \rightarrow 1$ boil down to the asymptotic causality condition $0<v_g^2<1$ for the MIS theory in the parameter range $\eta, \tau _{\pi }>0$.

For BDNK theory, the shear channel stability condition at  $\textbf{v} \rightarrow 1$ gives $\frac{\theta }{\eta }>1$, which is again the asymptotic causality condition $0<v_g^2<1$ where $v_g^2=\frac{\eta }{\theta }$. 
Next, for the BDNK sound channel, we attempt to solve the inequalities \eqref{BDNK-snd-stab1} and \eqref{BDNK-snd-stab2} served as stability criteria in a boosted frame. Stability inequality \eqref{BDNK-snd-stab1} can be recast as,
\begin{align}
\left\{\left(1/\textbf{v}^2\right) - x_1\right\} \left\{\left(1/\textbf{v}^2\right) - x_2\right\} > 0~,
\label{caus-BDNK-1}
\end{align}
where $x_1, x_2$ are the roots of the equation, 
\be
({\cal{E}}\theta) x^2-\frac{2}{3}{\cal{E}}(2\eta+\theta)x+\frac{1}{9}\theta({\cal{E}}-4\eta)=0 ~.
\label{caus-BDNK-2}
\ee
Inequality \eqref{caus-BDNK-1} has two possible solutions $x_1,x_2<\frac{1}{\textbf{v}^2}$ or $x_1,x_2>\frac{1}{\textbf{v}^2}$. Since $|\textbf{v}|$ ranges from 0 to 1 and hence $1/\textbf{v}^2$ ranges from 1 to $\infty$, the second solution turns out to the unphysical. The first and only physically acceptable solution then gives us the strictest bound $x_1,x_2<1$ corresponding to the limit $\textbf{v}\rightarrow 1$. Now, incorporation of the second stability inequality \eqref{BDNK-snd-stab2} restricts the allowed region to only positive values of $\mathcal{E}$ and $\theta$. This restriction (along with $\eta>0$) leads to a positive discriminant of \eqref{caus-BDNK-2}, which restricts both the roots of $x$ to be real, among which at least one root is always positive in our stable parameter space at  $\textbf{v} \rightarrow 1$. As it will be explicitly shown in the next section doing a large $k$ analysis of the theory that the quadratic equation satisfied by $v_g^2$ for the BDNK sound channel is exactly identical to \eqref{caus-BDNK-2}, the inequalities \eqref{BDNK-snd-stab1} and \eqref{BDNK-snd-stab2} condense down together to give $v_g^2<1$ with at least one $v_g^2>0$ that produces two subluminal propagating modes. So, our stability analysis at ultra-high boost independently identifies the causal parameter space of the MIS and BDNK theories, which exactly reproduces the results of asymptotic causality analysis for the respective theories without going to the large $k$ limit.

\section{Causality from large $k$ analysis}

Now, let us analyze the situation of causality in the high-$k$ regime itself and compare how accurately the subluminal parameter space has been predicted by stability analysis at ultra-high boost. At the large $k$ limit, an expansion of the form
$\omega = v_g k+\sum _{n=0}^{\infty }c_n k^{-n}$
is used \cite{Brito:2020nou} as a solution of the dispersion equation from which a polynomial over the asymptotic group velocity $v_g$ can be obtained. Next, we check the Schur stability of the polynomial \cite{Schur} to check if the roots of these equations are subluminal and, if they are, then how the parameter space is constrained by them. Any polynomial $P(z)$ of degree $d$ is called ``Schur stable" if its roots lie within a unit disc around the origin of the complex plane. This can be tested by introducing a M\"{o}bius transformation $w=(z+1)/(z-1)$, which maps the unit disc about the origin of the complex plane into the left half plane, i.e., $\text{Re} (w) <0$ if $|z|<1$. So, $P(z)$ will be Schur stable if and only if the transformed polynomial of the same degree $Q(w)=(w-1)^d P\left(\frac{w+1}{w-1}\right)$ is Hurwitz stable. This method is extremely efficient, especially in cases where a direct extraction of roots from the polynomial is too complicated.

For the shear channels, the Schur stability conditions that can give rise to subluminal, propagating modes are $\tau _{\pi }-\tilde{\eta }>0$ and $\tau _{\pi }+\tilde{\eta }>0$ for MIS and 
$\theta-\eta>0$ and $\theta+\eta>0$ for BDNK. In both cases, the first conditions are identically the stability conditions obtained at $\textbf{v} \rightarrow 1$ and the second conditions are obvious if the first ones are satisfied. For the propagating modes of MIS sound channel, the Schur stability conditions are given by $\tau _{\pi }-2\tilde{\eta }>0$ and $\tau _{\pi }+\tilde{\eta }>0$. Again, the first one is the $\textbf{v} \rightarrow 1$ stability criterion, and the rest is its obvious implication. So, we conclude that for both the shear channels and the MIS sound channel, the $\textbf{v} \rightarrow 1$ stability region exactly reproduces the causal parameter space. 

The situation in the BDNK sound channel is comparatively quite non-trivial. The $v_g^2$ values are to be extracted from the following quadratic polynomial with $z=v_g^2$,
\be
P(z)=({\mathcal{E}}\theta) z^2-\frac{2}{3}{\mathcal{E}}(\theta+2\eta)z+\frac{1}{9}\theta({\mathcal{E}}-4\eta)=0~,
\label{vg}
\ee
whose Schur stability needs to be checked to find the causal parameter space. Its M\"{o}bius transformation again turns out to be a quadratic polynomial, 
\begin{align}
 Q(w)=&\left(\frac{{\mathcal{E}}\theta}{3}-{\mathcal{E}}\eta-\frac{\eta\theta}{3}\right)w^2+\frac{2}{3}\theta\left(\eta+2{\mathcal{E}}\right)w\nonumber\\
      &+\left(\frac{4{\mathcal{E}}\theta}{3} +{\mathcal{E}}\eta-\frac{\eta\theta}{3}\right)=0~,
\label{Q}
\end{align}
whose Hurwitz stability requires all the three coefficients of Eq.\eqref{Q} to be of the same sign, either positive or negative (along with a positive discriminant of $P(z)$ to ensure that all the non-real roots of $v_g^2$ on the complex plane are excluded).

\begin{figure}[h!]
 \centering
    \includegraphics[width=0.42\textwidth]{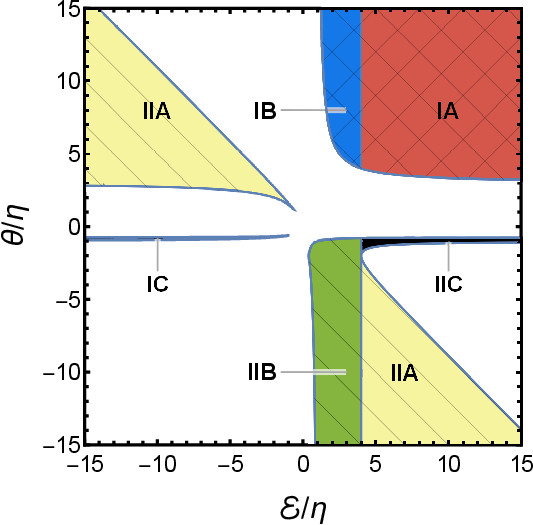}
    \caption{The subluminal parameter space for BDNK sound channel from Schur stability.}
    \label{fig:my_label2}
\end{figure}

In Fig.\ref{fig:my_label2}, the parameter space for which both the roots satisfy $\left| v_g^2 \right| <1$ are plotted for both the positive as well as negative conventions. The regions IA (red, crisscrossed), IB (blue, crisscrossed) and IC (black, solid-filled) are located within quadrants where both $\theta$ and $\mathcal{E}$ are of the same sign and indicate the regions of the parameter space where all the coefficients of \eqref{Q} are positive. The regions IIA (yellow, striped), IIB (green, striped) and IIC (black, solid-filled) are located within quadrants with $\theta$ and $\mathcal{E}$ of opposite signs and denote the convention where all coefficients of \eqref{Q} are negative. Together, all of these regions (IA-C, IIA-C) provide the full causal parameter space given by \eqref{vg}. Furthermore, the signs of the coefficients of \eqref{vg} indicate that the regions IC and IIC bounded by $\mathcal{E}>4\eta, \mathcal{E}<0, -2\eta < \theta <0$ give $-1 < v_g^2 <0$ for both roots and hence, fail to generate any propagating mode. The rest of the regions (IA-B, IIA-B) correspond to at least one $0<v_g^2 <1$ and hence at least two subluminal propagating modes. The regions IA and IIA cover the parameter space with the additional constraints $\mathcal{E}<0, \mathcal{E}>4\eta, \theta>0, \theta<-2\eta$, which give us both $v_g^2$ values between $0$ and $1$ and hence, four subluminal propagating modes. The remaining two regions, IB and IIB, belong to the parameter space constrained by $0< \mathcal{E} < 4\eta$, which corresponds to $-1<v_g^2<0$ for one root and $0<v_g^2<1$ for the other, indicating the presence of two non-propagating modes besides the existence of the two subluminal propagating modes.

Now comes a crucial identification; we observe that the causal parameter space in the first quadrant covered by the regions IA and IB together exactly agrees with the stable region at $\textbf{v} \rightarrow 1$ and hence, with the frame-invariantly stable parameter space as well. This can be readily checked by realizing that the Schur condition from \eqref{Q}, $-\frac{\eta \theta }{3}-{\mathcal{E}} \eta +\frac{{\mathcal{E}} \theta }{3}>0$ is exactly identical to the stability constraint \eqref{BDNK-snd-stab1} at $\textbf{v} \rightarrow 1$. The other two Schur conditions, $\theta  (\eta +2 {\mathcal{E}} )>0$ and $-\frac{\eta \theta }{3}+{\mathcal{E}} \eta +\frac{4 {\mathcal{E}} \theta }{3}>0$ along with a positive discriminant of \eqref{vg}, further restrict the region exclusively to within the $\theta>0, {\mathcal{E}}>0$ quadrant for propagating modes, which exactly resembles the role played by \eqref{BDNK-snd-stab2} with $\textbf{v} \rightarrow 1$ to define the stable parameter space. So, the entire causal parameter space obtained from the asymptotic equation \eqref{vg} (by Schur convention I, all coefficients $>0$) is fully identified by the stable region at ultra-high boost depicted in Fig.\ref{fig:my_label}. In this context, we refer to the results obtained in \cite{Hoult:2020eho}, where the large wave-number causality constraint is given solely by region IA with four subluminal propagating modes. The analysis there lacks the region IB where two subluminal propagating modes are present along with two non-propagating modes. We duly point out that this lacking region is stable in every reference frame (Fig.\ref{fig:my_label}), which invariably identifies this region to respect causality since covariant stability is possible only for causal systems \cite{Gavassino:2021owo, Gavassino:2023myj}. So, we conclude that, because of the complexity involved, it is indeed difficult to analytically extract the full causal parameter space from the large-$k$ dispersion polynomial. However, the method of stability analysis at $\textbf{v} \rightarrow 1$  presented in this work is much more effective in pointing out the full stable and causal parameter space unambiguously.

We finally point out that for regions IIA and IIB, where $\theta$ and ${\mathcal{E}}$ are of opposite signs, the system is unstable in all reference frames. As mentioned in the stability arguments of \cite{Gavassino:2021owo}, there could be other regions of the parameter space like IIA and IIB, where causality holds, but the system is invariantly unstable in all reference frames. The stability criteria at ultra-high boost strictly give us the parameter space where these two theories are causal as well as stable in all reference frames.

\section{Conclusion}

We have shown here, for the first time, for two well-known stable-causal hydrodynamic theories, viz. MIS and BDNK, an alternate way to derive the region of parameter space over which the theories are 
frame-invariantly stable at leading order in $k$ and necessarily causal. Despite inherent differences in their construction, our analysis reveals that linearized stability analysis at ultra-high boost accurately 
leads us to the results of asymptotic causality conditions under which both the theories are frame-invariantly stable, without going to the large-$k$ limit. Since the whole analysis is performed at a low-$k$ limit, this approach liberates us from going to a non-perturbative high-$k$ regime that seems outside the domain of validity of a low-energy effective theory like relativistic hydrodynamics. Moreover, in the presence of technical non-trivialities in solving the asymptotic causality equations, our method of stability check at $\textbf{v}\rightarrow 1$ is more effective and simpler in detecting the causal parameter space.
Although the current analysis has been carried out for a conformal, chargeless system, the results presented here do not lack in generality. In \cite{Biswas:2022hiv}, a coarse-grained derivation of a 
non-conformal, charged, stable-causal first-order theory indeed shows that the monotonically decreasing stable parameter space becomes the strictest bound for $\textbf{v}\rightarrow 1$ which singularly gives the causal parameter space as well. 

The findings presented here heavily depend upon the monotonic behavior of the stable parameter space as a function of $\textbf{v}$. The monotonic behavior that exists for these two most well-known stable-causal theories doesn't hold for the relativistic first-order Navier-Stokes theory. This indicates that this feature could be an important signature for pathology-free hydrodynamic theories. Further, the prediction of high-$k$ results from the low-$k$ domain using ultra-high boost, as observed here, indicates some possible connection between the two limiting $k$-regimes of the theories, which requires further investigation. In Appendix-A we have derived our results for a more general class of hydrodynamic problems and provided intuitive arguments in support of the current outcome.
The causality criteria considered here are asymptotic causality criteria which are necessary but not sufficient conditions \cite{Kundt}. A more rigorous study of causality requires a study of characteristics \cite{Bemfica:2019cop,Bemfica:2020xym}, which will be explored in our future endeavors. Further, studies related to higher order theories in a general hydrodynamic frame \cite{Noronha:2021syv} and comparative studies between the BDNK and MIS theories \cite{Das:2020fnr} are also in line to be explored in future. Lastly, observing the effects of the bounds provided in this study to the fluid parameter space on the numerical hydro simulations \cite{Chiu:2021muk,Plumberg:2021bme,Pandya:2021ief} is another direction to explore.

\section*{Appendix A - Causality criteria from near-luminal stability for a general hydrodynamic theory}
\subsection{Monotonic behavior of stable parameter space with boost velocity}
In the current analysis, the conservation equations (giving rise to hydrodynamic evolution equations) are linearized for small perturbations of fluid variables around their hydrostatic equilibrium. The method gives the dispersion polynomial in the frequency $(\omega,\k)$ plane as $F(\omega,\k)=0$, whose solution provides the dispersion relation $\omega=\omega(\k)$ that is required for the stability analysis. Here, we are deriving our results for a general hydrodynamic dispersion polynomial (irrespective of shear or sound channel), which obeys just two assumptions guided by generic physics requirements. The assumptions are motivated by the conservation rules (of the number of fluid modes) and the symmetry requirements and do not compromise the generality of our method.

{\bf{Assumption 1 :}}
The total power of any term that contains $\k$ (it can be a term that contains only $\k$ or an admixture of $\omega$ and $\k$) must not exceed the largest power of a pure $\omega$ term. Following this criteria, a most general dispersion polynomial must obey,
\begin{equation}
 {\cal{O}}_{\omega}[F(\omega,\k\neq0)]= {\cal{O}}_{\vert\k\vert}[F(\omega=a\vert\k\vert,\k=\bf{b}\vert{\k}\vert]~,
 \label{genpoly}
\end{equation}
with $a$ as a nonzero real scalar constant, $\bf{b}$ as a real unit vector and ${\cal{O}}_x$ denoting the order of the polynomial in the variable $x$.

In Ref.\cite{Hoult:2023clg}, Eq.\eqref{genpoly} has been mentioned as a condition for causality. We are justifying this assumption from the point of Lorentz invariance of the number of modes in a theory. If the right-hand side of \eqref{genpoly} has a larger order than the left-hand side, then a Lorentz boost of the background fluid with a velocity $\v$ always produces spurious modes, modes that never appeared in the local rest frame analysis \cite{Denicol:2008ha,Mitra:2021ubx}. Given that the number of modes changes with the arbitrary choice of equilibrium state, it is indicative that the equations of motion that lead to such a polynomial cannot constitute a viable theory
of viscous hydrodynamics. Moreover, the solution of these new modes will be inversely proportional to some powers of $\v$ (in the boosted frame the polynomial variable changes from $\omega$ to $\v\omega$), that diverges as $\v\rightarrow 0$ and hence are unphysical. With this chain of arguments, below we are writing the most general form of the dispersion polynomial (of order $M$) for any arbitrary hydrodynamic theory in the local rest frame of the fluid:
\be
 a_{M}\om^{M}+a_{M-1}\om^{M-1}+\cdots+a_2\om^2\nn
 +a_1\om+a_0=0~,
 \label{poly}
\ee
with,
\begin{align}
&a_0=a_0^0+a_0^1 k+\cdots+a_0^{M-2}k^{M-2}+a_0^{M-1}k^{M-1}+a_0^{M}k^{M},\nn
&a_1=a_1^0+a_1^1 k+\cdots+a_1^{M-2}k^{M-2}+a_1^{M-1}k^{M-1}~,\nn
&a_2=a_2^0+a_2^1 k+\cdots+a_2^{M-2}k^{M-2}~,\nn
&\vdots\nn
&a_{M-2}=a_{M-2}^0+a_{M-2}^1 k+a_{M-2}^2 k^2~,\nn
&a_{M-1}=a_{M-1}^0+a_{M-1}^1 k~,\nn
&a_M=a_M^0~,
\label{polycoeff}
\end{align}
which in a consolidated form can be written as,
\begin{equation}
    \begin{split}
        \sum_{n=0}^M a_n (k)\omega^n &=0 ,~~~~
        a_n(k) = \sum_{m=0}^{M-n} a_n^m k^m~.
    \end{split}
\end{equation}
The coefficients $a_n^m$ (the subscript $n$ denotes the power of $\om$ and the superscript $m$ denotes the power of $k$ of the term it is associated with) are functions of transport coefficients of the underlying coarse-grained system that set the parameter space of the theory. We are putting no constraint on the $a_n^m$ values. They can be both real and imaginary and can have positive or negative values or even become zero depending upon the construction of a particular hydrodynamic theory.

Our next step is to boost Eq.\eqref{poly} with velocity $\v$ and extract the stability criteria of that boosted polynomial at the spatial homogeneous limit   ($k\rightarrow 0$). At $k\rightarrow 0$, the boosted form of Eq.\eqref{poly} becomes,
\begin{align}
&\left(\gm\om\right)^{M}\bigg[a_M^0(-\v)^0+a_{M-1}^1(-\v)^1+a_{M-2}^2(-\v)^2+\cdots\nn
&~~~~~~~~+a_{2}^{M-2}(-\v)^{M-2}+a_{1}^{M-1}(-\v)^{M-1}+a_{0}^{M}(-\v)^{M}\bigg]\nn
&+\left(\gm\om\right)^{M-1}\bigg[a^0_{M-1}(-\v)^0+a^1_{M-2}(-\v)^1+\cdots\nn
&~~~~~~~~~~~~~+a_1^{M-2}(-\v)^{M-2}+a_0^{M-1}(-\v)^{M-1}\bigg]+\cdots\nn
&+\left(\gm\om\right)^{1}\bigg[a_1^0(-\v)^0+a_0^1(-\v)^1\bigg]+\left(\gm\om\right)^{0}\bigg[a_0^0(-\v)^0\bigg]=0~,
\label{polyboost}
\end{align}
with $\gm=1/\sqrt{1-\v^2}$. Eq.\eqref{polyboost}
can again be expressed in a general form as,
\begin{equation}
   \sum_{n=0}^M A_n (\gamma \omega)^n =0, ~~~~ A_n = \sum_{m=0}^n a_m^{n-m} (-\v)^{n-m}~.
\end{equation}
Since an analytical solution of Eq.\eqref{polyboost} is beyond the scope, in order to check its stability we take recourse of Routh-Hurtwitz (R-H) stability test \cite{Korn}. The stability condition requires the elements belonging to the first column of the Routh array (includes the coefficients of $(\gm\om)^M$, $(\gm\om)^{M-1}$ and determinants involving other coefficients of \eqref{polyboost})
to be of identical sign, either positive or negative. This leads us to $M+1$ number of inequalities which say that, in order to have a stable theory, all these elements are either greater or lesser than zero. So if these elements are expressed as $f_i\left(\{a_n^m\},\v\right)$, for all roots of $\omega$ to be stable, we must have either
\begin{equation}
    f_i\left(\{a_n^m\},\v\right) > / < 0~,
\end{equation}
for all $i \in \{1, M+1\}$.
At this point, we state our second assumption.
\newline
\newline
{\bf{Assumption 2 :}}
The local rest frame dispersion polynomial \eqref{poly} only allows even power of $\abs{\k}=\sqrt{\k^2}$, making it $F(\om,\k^2)=0$, i.e, the coefficients $a_n^m$ with an odd $m$ are zero \cite{Grozdanov:2019uhi}. This can be simply understood from the fact that $\k$ being a vector, only the powers of $\k^2$ are allowed in the scalar dispersion polynomial \eqref{poly}. As a consequence, the boosted polynomial \eqref{polyboost} contains only even power of $\abs{\v}=\sqrt{\v^2}$ (also required since $\v$ is a vector as well).

These two assumptions lead to the fact that the R-H stability criteria of \eqref{polyboost} boil down to a set of inequalities where a power series over $\v^2$ is greater or lesser than zero. To demonstrate the situation we are writing here the condition over the first element of the first column of the Routh array,
\begin{align}
 a_M^0(\v)^0+a_{M-2}^2(\v)^2+\cdots
+a_{2}^{M-2}(\v)^{M-2}+a_{0}^{M}(\v)^{M} > 0~.
\label{inequality}
\end{align}
Here, $M$ is considered to be even (odd $M$ conditions can be similarly extracted where the power of the last term would be $M-1$) and we illustrate the result for the ``all positive" possibility.  Now, the left-hand side of inequality \eqref{inequality} can be decomposed as,
\begin{align}
 (\v^2-x_1)(\v^2-x_2)\cdots(\v^2-x_{M/2})>0~,
\label{root1}
 \end{align}
where $x_l$ are the roots of the polynomial,
\begin{align}
 a_M^0+a_{M-2}^2x+\cdots
+a_{2}^{M-2}x^{M/2-1}+a_{0}^{M}x^{M/2} = 0~,
\label{root2}
\end{align}
and are functions of the $a_n^m$ coefficients only (i.e., $x_l \equiv x_l(a_n^m)$), which are again functions of the transport coefficients of the system.
So, to hold inequality \eqref{root1}, each factor $(\v^2-x_l)$ has to be positive or negative accordingly. So finally the R-H criteria boil down to a set of inequalities such that,
\begin{equation}
(\v^2-x_l)>/< 0~~~~=>~~~~x_l(a_n^m)>/< \v^2~.
\label{cond1}
\end{equation}
So, from \eqref{cond1} we can see that the stability criteria of any theory reduces to a set of inequalities where a function of the fluid parameters is greater or lesser than $\v^2$. Clearly, this indicates a monotonic behavior of the parameter space on $\v^2$, and consequently, at spatial homogeneous limit $(k\rightarrow 0)$, the stable parameter space must monotonically decrease from $\v=0$ to $1$ or from $\v\rightarrow 1$ to $0$ respectively. So, if we follow the `greater than' possibility ($x_l(a_n^m)> \v^2$) of \eqref{cond1}, the stability region of parameter space for $\v\rightarrow 1$ includes the same for any lower value of $\v$ turning the stability condition at $\v\rightarrow 1$ a necessary and sufficient condition for stability to hold at the spatially homogeneous limit for all possible boost velocities $0\leq\v<1$. Conversely, following the `lesser than' possibility, the direction of monotonicity reverses.

Now, the sign of the inequality in \eqref{cond1} (that leads to the direction of monotonicity) suffers from ambiguity. The reason is that since Eq.\eqref{poly} describes the dispersion polynomial of a possible most general theory, the signs of the $a_n^m$ coefficients are completely unknown and arbitrary. To resolve this ambiguity, we investigate Eq.\eqref{polyboost} at different boost velocities and provide the following line of arguments.

At $v=0$ we observe that for each $n$, only the coefficients $a_n^m$ with $m=0$ are contributing to the stability analysis. For a non-zero value of $\v$, all the $a_n^m$ coefficients with even $m$ are contributing. If we have a look at Eq.\eqref{polycoeff}, we can see that the stability conditions at non-zero $\v$ constrain a much larger number of elements of the parameter space, making the system of inequalities more restrictive than the ones at $\v=0$. In other words, the conditions at $\v\neq 0$ lay a stricter bound on the entire parameter space than those at $\v=0$. So, it is indicative that the monotonicity over $\v^2$ that has been discussed so far is uniformly restricting the parameter space from $\v=0$ to $\v\rightarrow 1$. This turns the parameter space, which is stable at near-luminal boost velocity, a necessary and sufficient region for frame-invariant
stability to hold (at the spatially homogeneous limit),
and consequently, identify the causal parameter space as well \cite{Gavassino:2021owo}.

In support of the above discussion, here we are writing the polynomial equation for asymptotic group velocity $v_g$ at $k\rightarrow\infty$ resulting from \eqref{poly} for even $M$ :
\be
 a_M^0(v_g^2)^{\frac{M}{2}}+a_{M-2}^2(v_g^2)^{\frac{M}{2}-1}+\cdots+a_2^{M-2}v_g^2+a_0^M=0~.
\label{asymp}
 \ee
In order to have a causal, propagating mode, \eqref{asymp} must have real, positive, subluminal roots of $v_g^2$, which are the functions of the $a_{n}^{m}$ coefficients of \eqref{asymp}. From Eq.\eqref{polycoeff}, we see that these $a_n^m$ are the coefficients of the largest $k$ power for each $a_n$ term with $n$ even. Clearly, the conditions for subluminal roots will involve constraints on these coefficients. Here, we see that stability conditions for $\v=0$ only include the $a_M^0$ among these coefficients and can not able to identify the causal parameter space because of this nominal overlap. On the other hand,  the stability constraints with nonzero $\v$ include all the coefficients of Eq.\eqref{asymp}. So, the monotonicity over $\v^2$ leaves us with the choice that stability at $\v\rightarrow 1$ demarcates the causal parameter space.
\subsection{Connection between stable parameter space at $k\to 0$, $\v\to 1$ and causal parameter space at large $k$ }
A mathematical explanation regarding this connection can be followed here. For that, say for an even $M$ case we divide Eq.\eqref{inequality} by $(\v^2)^{M/2}$. Being a positive quantity, it will not alter the sign of inequality and converts \eqref{inequality} into,
\begin{align}
&a_M^0\left(\frac{1}{\v^2}\right)^\frac{M}{2}+a_{M-2}^2\left(\frac{1}{\v^2}\right)^{\frac{M}{2}-1}+\cdots\nn
&~~~~~~~~~~~~~~~+a_{2}^{M-2}\left(\frac{1}{\v^2}\right)+a_{0}^{M} > 0~,
\label{inequality2}
\end{align}
which can be decomposed as,
\begin{align}
 \left(\frac{1}{\v^2}-y_1\right)\left(\frac{1}{\v^2}-y_2\right)\cdots\left(\frac{1}{\v^2}-y_{M/2}\right)>0~,
 \label{inequality22}
\end{align}
with $y_l$ being roots of,
\begin{align}
a_M^0y^\frac{M}{2}+a_{M-2}^2y^{\frac{M}{2}-1}+\cdots
+a_{2}^{M-2}y+a_{0}^{M} = 0~.
\label{inequality3}
\end{align}
Now, in order to hold inequality \eqref{inequality22}, each bracketed quantity on the left-hand side has to be individually positive or negative.
The only physical choice is the positive convention, which for each $y_l$ leads to,
\begin{align}
 \left(\frac{1}{\v^2}-y_l\right)>0~~,~~~~y_l<\frac{1}{\v^2}~,
 \label{inequality4}
\end{align}
which gives the strictest bound at $\v\rightarrow 1$ such that $y_l < 1$. Here we make an important observation. We notice that Eq.\eqref{inequality3} and the polynomial for asymptotic group velocity \eqref{asymp} are identical. Consequently, the $y_l$ are the solutions for $v_g^2$ itself. So from \eqref{inequality4}, we can see that the stability conditions at $\v\rightarrow 1$ are indeed related to the causality criteria of the theory ($v_g^2<1$). It is to be noted here that \eqref{inequality} is not the only stability condition (it is the first one of them; there are $M$ more). In the theories that we have studied in our work - MIS and BDNK - the other conditions basically set the convention for the direction of inequalities that cancels any choice of $y_l$ other than \eqref{inequality4}. Nevertheless, the structural similarity of \eqref{asymp} and \eqref{inequality2} is enough to indicate the connection between the near luminal ($\v\rightarrow 1$) stability conditions at the spatial homogeneous limit ($k\rightarrow 0$) and the causality criteria predicted at the asymptotic causality limit ($k\rightarrow \infty$).

\section*{Acknowledgements} 

We duly acknowledge Sayantani Bhattacharyya and Victor Roy for useful discussions, valuable inputs and critical reading of the manuscript. We also acknowledge Anirban Dinda, Archisman Bhattacharjee and Najmul Haque for their valuable discussions. The authors acknowledge financial support from the Department of Atomic Energy, India.

\end{document}

\title{Causality Criteria from Stability Analysis at Ultra-High Boost - Supplementary material}

\author{Shuvayu Roy}
\email{shuvayu.roy@niser.ac.in}
\affiliation{School of Physical Sciences, National Institute of Science Education and Research, An OCC of Homi Bhabha National Institute, Jatani-752050, India.}

\author{Sukanya Mitra}
\email{sukanya.mitra@niser.ac.in}
\affiliation{School of Physical Sciences, National Institute of Science Education and Research, An OCC of Homi Bhabha National Institute, Jatani-752050, India.}

\begin{abstract}
 We have provided here the detailed analytical expressions of the dispersion polynomials for MIS and BDNK theory for both the shear and sound channels. These dispersion relations have been derived 
 in a boosted reference frame with arbitrary background velocity {\textbf{v}}, and for all general values of wave number $k$. From these dispersion relations, the frequency modes are
 extracted that have been used in the main text for the stability and causality analysis.
\end{abstract}

\maketitle

The MIS shear channel dispersion relation in a boosted frame with background velocity {\textbf{v}} and  for all values of $k$ is given by,
\begin{align}
&\gamma\left(\tau_{\pi}-\tilde{\eta}\textbf{v}^2\right)\omega^2+\left[i-2\gamma\textbf{v}(\tau_{\pi}-\tilde{\eta})k\right]\omega\nonumber\\
& + \left[-i\textbf{v}k+\gamma\left(\tau_{\pi}\textbf{v}^2-\tilde{\eta}\right)k^2\right]=0~.
\end{align}
For MIS sound channel, the dispersion relation reads,
\begin{align}
 &A_3^0\omega^3+\left(A_2^0+A_2^1k\right)\omega^2+\left(A_1^1k+A_1^2k^2\right)\omega\nonumber\\
 &+\left(A_0^2k^2+A_0^3k^3\right)=0~,
 \label{MIS-shear}
\end{align}
with,
\begin{align}
 A_3^0&=\tau_{\pi}-\left(\frac{4\tilde{\eta}}{3}+\frac{\tau_{\pi}}{3}\right)\textbf{v}^2~,\nonumber\\
 A_2^0&=i\frac{1}{\gamma}\left(1-\frac{\textbf{v}^2}{3}\right)~,\nonumber\\
 A_2^1&=\left(\frac{8\tilde{\eta}}{3}-\frac{7\tau_{\pi}}{3}\right)\textbf{v}+\left(\frac{4\tilde{\eta}}{3}+\frac{\tau_{\pi}}{3}\right)\textbf{v}^3~,\nonumber\\
 A_1^1&=-i\frac{1}{\gamma}\frac{4}{3}\textbf{v}~,\nonumber\\
 A_1^2&=-\left(\frac{4\tilde{\eta}}{3}+\frac{\tau_{\pi}}{3}\right)+\left(\frac{7\tau_{\pi}}{3}-\frac{8\tilde{\eta}}{3}\right)\textbf{v}^2~,\nonumber\\
 A_0^2&=i\frac{1}{\gamma}\left(\textbf{v}^2-\frac{1}{3}\right)~,\nonumber\\
 A_0^3&=-\tau_{\pi}\textbf{v}^3+\left(\frac{4\tilde{\eta}}{3}+\frac{\tau_{\pi}}{3}\right)\textbf{v}~.
 \label{MIS-sound}
\end{align}
The dispersion relation for BDNK shear channel is identical to that of MIS shear channel \eqref{MIS-shear}, with the replacement, $\tau_{\pi}=\tilde{\theta}$.

The dispersion relation for BDNK sound channel is given below,
\begin{align}
 &C_4^0 \omega^4 + \left(C_3^0+C_3^1 k\right)\omega^3 + \left(C_2^0+C_2^1k+C_2^2k^2\right)\omega^2 +\nonumber\\
 &\left(C_1^1k+C_1^2k^2+C_1^3k^3\right)\omega + \left(C_0^2k^2+C_0^3k^3+C_0^4k^4\right) =0~,
\end{align}
with,
\begin{align}
 C_4^0& =\gamma^4\left[\tilde{\mathcal{E}}\tilde{\theta}-\frac{2}{3}\tilde{\mathcal{E}}\left(2\tilde{\eta}+\tilde{\theta}\right)\textbf{v}^2
 +\frac{1}{9}\tilde{\theta}\left(\tilde{\mathcal{E}}-4\tilde{\eta}\right)\textbf{v}^4\right]~,\nonumber\\
 C_3^0& = i\gamma^3\left[\left(\tilde{\mathcal{E}}+\tilde{\theta}\right)-\frac{1}{3}\left(\tilde{\mathcal{E}}+\tilde{\theta}+4\tilde{\eta}\right)\textbf{v}^2\right]~,\nonumber\\
 C_3^1&= \gamma^4\left[\frac{8}{3}\tilde{\mathcal{E}}\left(\tilde{\eta}-\tilde{\theta}\right)\textbf{v} + \frac{8}{3}\left(\tilde{\mathcal{E}}\tilde{\eta}+\frac{2}{3}\tilde{\eta}\tilde{\theta}+
 \frac{1}{3}\tilde{\mathcal{E}}\tilde{\theta}\right)\textbf{v}^3\right]~,\nonumber\\
 C_2^0&=\gamma^2\left(\frac{\textbf{v}^2}{3}-1\right)~,\nonumber\\
 C_2^1&=i\gamma^3\left[\left\{-\frac{7}{3}\left(\tilde{\mathcal{E}}+\tilde{\theta}\right)+\frac{8}{3}\tilde{\eta}\right\}\textbf{v}
 +\frac{1}{3}\left(\tilde{\theta}+\tilde{\mathcal{E}}+4\tilde{\eta}\right)\textbf{v}^3\right]~,\nonumber\\
 C_2^2&=\gamma^4\big[-\frac{2}{3}\tilde{\mathcal{E}}\left(2\tilde{\eta}+\tilde{\theta}\right)+\left(-\frac{16}{3}\tilde{\eta}\tilde{\mathcal{E}}-\frac{8}{3}\tilde{\eta}\tilde{\theta}+4\tilde{\mathcal{E}}\tilde{\theta}\right)\textbf{v}^2\nonumber\\
      &~~~~~~~~~-\frac{2}{3}\tilde{\mathcal{E}}\left(2\tilde{\eta}+\tilde{\theta}\right){\textbf{v}^4}\big]~, \nonumber\\     
 C_1^1&=\gamma^2\frac{4}{3}\textbf{v}~,\nonumber\\
 C_1^2&=i\gamma^3\left[-\frac{1}{3}\left(\tilde{\theta}+\tilde{\mathcal{E}}+4\tilde{\eta}\right)+\left\{\frac{7}{3}\left(\tilde{\mathcal{E}}+\tilde{\theta}\right)-\frac{8}{3}\tilde{\eta}\right\}\textbf{v}^2 \right]~,\nonumber\\
 C_1^3&=\gamma^4\left[\frac{8}{3}\left(\tilde{\mathcal{E}}\tilde{\eta}+\frac{2}{3}\tilde{\eta}\tilde{\theta}+
 \frac{1}{3}\tilde{\mathcal{E}}\tilde{\theta}\right)\textbf{v}
 +\frac{8}{3}\tilde{\mathcal{E}}\left(\tilde{\eta}-\tilde{\theta}\right)\textbf{v}^3 \right]~,\nonumber\\
 C_0^2&=\gamma^2\left(\frac{1}{3}-\textbf{v}^2\right)~,\nonumber\\
 C_0^3&=i\gamma^3\left[\frac{1}{3}\left(\tilde{\mathcal{E}}+\tilde{\theta}+4\tilde{\eta}\right)\textbf{v}-\left(\tilde{\mathcal{E}}+\tilde{\theta}\right)\textbf{v}^3\right]~,\nonumber\\
 C_0^4&=\gamma^4\left[\frac{1}{9}\tilde{\theta}\left(\tilde{\mathcal{E}}-4\tilde{\eta}\right)-\frac{2}{3}\tilde{\mathcal{E}}\left(2\tilde{\eta}+\tilde{\theta}\right)\textbf{v}^2
  +\tilde{\mathcal{E}}\tilde{\theta}\textbf{v}^4\right]~.
 \end{align}